\documentclass[twocolumn,pra]{revtex4}

\usepackage{amsmath}
\usepackage{amsthm}
\usepackage{algorithm,algpseudocode}
\usepackage{algorithmicx}
\usepackage{color}
\usepackage{float}
\usepackage{amssymb}
\usepackage{stmaryrd}
\usepackage{graphicx}
\usepackage{epstopdf}

\algrenewcommand\algorithmicrequire{\textbf{Given:}}
\algrenewcommand\algorithmicensure{\textbf{Return:}}

\definecolor{red}{rgb}{1,0.,0}
\definecolor{white}{rgb}{1.,1.,1.}

\global\long\def\comsign#1{\llbracket#1\rrbracket}

\global\long\def\comsign#1{\llbracket#1\rrbracket}
\newtheorem{theorem}{Theorem}
\newtheorem{lemma}{Lemma}
\newtheorem{example}{Example}

\newcommand{\bs}[1]{\boldsymbol{#1}}

\input{Qcircuit}   

\date{12 May 2017}

\begin{document}

\newcommand{\ceil}[1]{\left\lceil #1 \right\rceil}

\title{Unbiased Simulation of Near-Clifford Quantum Circuits}
\author{Ryan S.~Bennink}
\author{Erik M.~Ferragut}
\author{Travis S. Humble}\email[]{humblets@ornl.gov}
\author{Jason A.~Laska}
\author{James J.~Nutaro}
\author{Mark G.~Pleszkoch}
\author{Raphael C.~Pooser}
\affiliation{Quantum Computing Institute, Oak Ridge National Laboratory, Oak Ridge, TN, USA}

\begin{abstract}
Modeling and simulation is essential for predicting and verifying the behavior of fabricated quantum circuits, but existing simulation methods are either impractically costly or require an unrealistic simplification of error processes. We present a method of simulating noisy Clifford circuits that is both accurate and practical in experimentally relevant regimes. In particular, the cost is weakly exponential in the size and the degree of non-Cliffordness of the circuit. Our approach is based on the construction of exact representations of quantum channels as quasiprobability distributions over stabilizer operations, which are then sampled, simulated, and weighted to yield unbiased statistical estimates of circuit outputs and other observables.  As a demonstration of these techniques we simulate a Steane [[7,1,3]]-encoded logical operation with non-Clifford errors and compute its fault tolerance error threshold. We expect that the method presented here will enable studies of much larger and more realistic quantum circuits than was previously possible. 
\end{abstract}

\maketitle

\section{Introduction}
Modeling and simulation of quantum circuits is indispensable for evaluating the potential of quantum computing devices and guiding their development. State-of-the-art approaches to simulation are limited to circuits involving either a relatively small number of qubits or a highly restricted set of gates (e.g. error-free Clifford gates). As the dynamics of a general quantum physical system are in the complexity class BQP, there is almost certainly no scalable method of simulating universal quantum circuits on conventional computers. This is an important issue for the ongoing development of quantum computing technology, as the utility of modeling and simulation is likely to be severely limited when experimental systems reach sizes significant for quantum computing.
\par
Although the cost of simulating quantum circuits on conventional computers is prohibitive in general, there are important exceptions. In particular, Gottesman and Knill showed that stabilizer circuits---circuits involving only Clifford operations and Pauli measurements---can be simulated efficiently on classical computers via stabilizer propagation \cite{gottesman1998heisenberg}. While these circuits are not universal for quantum computing, they are significant for having a prominent role in quantum error correction \cite{Gaitan2008}. Very large instances of these circuits can be simulated due to the efficient scaling of stabilizer propagation in both memory and run time, respectively, $O(n^2)$ and $O(k n^2)$ for $n$ qubits and $k$ gates.
\par
Physical realizations of stabilizer circuits often incur errors as a result of imperfections in gate operation and uncontrolled interactions with the environment. Stabilizer propagation can be extended with Monte Carlo techniques to account for errors modeled as probabilistic mixtures of stabilizer operations. In this approach, the behavior of the circuit is taken to be the average of many realizations of the circuit with random stabilizer errors. However, many common error processes including amplitude damping, over/under-rotation, and misaligned rotation cannot be represented exactly as probabilistic mixtures of stabilizer operations.  Several groups have investigated the honesty and accuracy of approximating such error processes as probabilistic mixtures of stabilizer operations \cite{Magesan2013, Gutierrez2015}. In some of the cases studied the stochastic approximation was found to have little impact on simulation results \cite{Puzzuoli2014, Gutierrez2015}, while in other cases it led to predictions of circuit failure rates that were in error by substantial factors \cite{Gutierrez2016}.
\par
We address the problem of simulating stabilizer circuits with non-stabilizer errors by representing non-stabilizer errors as \emph{quasiprobability} distributions over stabilizer operations. As we show in the Appendix, relaxing the usual constraint that probabilities be non-negative permits an exact representation of any quantum process (including non-unitary processes, non-unital processes, and general measurements) as a weighted sum of stabilizer operations. Our simulation method begins by representing each gate in the circuit in such a way. The circuit behavior is then estimated by a weighted average over realizations of the circuit with stabilizer errors randomly drawn from each gate's quasiprobability distribution. Thus our method is a direct generalization of existing stochastic stabilizer propagation methods to allow simulation of arbitrary gates and errors. The price to be paid for full generality is that the cost of simulation is generally exponential in the circuit size. But critically, the coefficient of the exponent depends on the degree of ``non-Cliffordness'' of the circuit as measured by the negativity of the gates' quasiprobability representations. For fault tolerance circuits in experimentally relevant regimes the exponential growth is very weak, so that simulations of circuits with hundreds of qubits are feasible.  
\par
Our approach complements the recent work of Pashayan et al., who formalized the role of quasiprobabilities in simulating quantum circuits \cite{Pashayan2015}. Their work identified the significance of negativity in a distribution for determining the cost of simulation. This can be related to the fact that interference between probabilities (i.e.\ negativity) is a necessary condition for quantum computational speedup \cite{Stahlke2014}. Others have shown that negativity plays an important role in determining the complexity of simulations involving quasiprobability distributions over circuit input states \cite{1367-2630-14-11-113011}. While our approach is similar in spirit to Pashayan et al., we have formulated circuit simulation specifically in terms of the qubit stabilizer formalism and stabilizer circuits.
\par
The paper is organized as follows: Sec.~\ref{sec:rep} briefly reviews the existing theory of stabilizer circuit simulation before presenting a representation of arbitrary quantum channels as quasiprobability distributions over stabilizer channels. In Sec.~\ref{sec:mc}, we discuss the scaling of Monte Carlo simulations using the quasiprobability representation and describe its implementation in Sec.~\ref{sec:ex}. We then demonstrate our method by simulating a Steane [[7,1,3]]-encoded logical operation with amplitude damping errors and computing the logical error probability. We offer final remarks in Sec.~\ref{sec:conc}.
\section{Stabilizer Channel Decomposition}
\label{sec:rep}
Our simulation method is based on the exact representation of quantum channels as quasiprobability distributions over stabilizer channels.  Quantum channels represent a very broad class of physical processes including ideal circuit operations and error processes as well as measurements \cite{nielsen2010quantum}.  Mathematically, a quantum channel is a linear mapping on the space of quantum states (the set of positive-semidefinite trace-1 operators).  We denote the action of a quantum channel $\mathbf{A}$ on a state $\rho$ as either $\mathbf{A}(\rho)$ or equivalently $\mathbf{A}\vec{\rho}$, where $\vec{\rho}$ denotes the vectorization of $\rho$. Channels should be contrasted with operators, which may act on states from the left or right. The channel corresponding to conjugation by an operator $A$ ($\rho \to A \rho A^\dagger$) will be denoted by the same symbol in bold ($\mathbf{A}$).
\par
The stabilizer formalism concerns a particularly useful set of quantum states and operations that originally arose in studies of error correction in circuits with Pauli errors~\cite{gottesman1998heisenberg}. 
An $n$-qubit stabilizer state is a fixed point of $n$ independent $n$-qubit Pauli operators. Stabilizer states include the eigenstates of Pauli operators, as well as many highly entangled states useful for quantum computing and error correction.
We call any quantum channel that maps stabilizer states to stabilizer states a \emph{stabilizer channel}. The set of stabilizer channels includes Clifford channels (channels corresponding to the unitary operations known as Clifford operations) as well as non-unitary channels, i.e. channels that are many-to-one on the space of stabilizer states. 
 \par
The great utility of the stabilizer formalism in the context of quantum error correction has prompted a number of authors to consider how it might be leveraged in broader contexts. Aaronson and Gottesman showed how arbitrary pure states and arbitrary unitary operators could be expressed as complex superpositions of stabilizer states and Pauli operators, respectively, thereby admitting simulation of arbitrary circuits ~\cite{Aaronson2004}. They provided a worst-case analysis in which the cost of simulation increases by a factor of $16^m$ for each non-Clifford $m$-qubit gate in the circuit. Garc\'ia and Markov introduced the concept of stabilizer frames to represent arbitrary pure states more compactly and demonstrated their effectiveness compared to existing simulation methods~\cite{garcia2015simulation}.
Yoder developed a similar representation for arbitrary states (including mixed states), consisting of a density operator in a stabilizer-defined basis~\cite{Yoder2012}. Most recently, Brayvi and Gosset developed techniques involving linear combinations of stabilizer states to simulate circuits involving Clifford gates and limited numbers of $T$ gates~\cite{bravyi2016improved}. 
\par
Others have taken the approach of simulating circuits with gates expressed as probabilistic mixtures (as opposed to complex superpositions) of stabilizer operations~\cite{Gutierrez2013, Magesan2013,Puzzuoli2014,Gutierrez2015}. Monte Carlo simulation of these circuits has a cost proportional to the cost of stabilizer propagation, which is between linear and quadratic in the number of qubits~\cite{Aaronson2004}. However, important quantum operations such as the $T$ gate, as well as realistic errors such as amplitude damping and small unitary errors, cannot be exactly represented as probabilistic mixtures of stabilizer operations, thereby limiting the applicability of that approach.
\par
In this work we generalize the stochastic approach by relaxing the constraint that the weights of the stabilizer operations be non-negative. The following lemma is proved in Appendix~\ref{sec:decomp}:
\begin{lemma}
\label{thm:channel-representation}
Any trace-preserving quantum channel on $n$ qubits can be expressed as a real linear combination of Clifford channels and Pauli reset channels on those qubits. If the channel is unital, Clifford channels suffice.
\end{lemma}
\noindent Here a {\em Pauli reset channel} refers to the process of measuring a Pauli operator and then conditionally performing a Clifford operation to place the qubit into a prescribed eigenstate of the Pauli operator \cite{Gutierrez2013}. An example is the case of initializing a qubit to the $\ket{0}$ state by measuring the $Z$ eigenvalue and performing an $X$ operation if the $-1$ eigenvalue was obtained. While there are more complicated stabilizer channels that could also be considered---for instance channels corresponding to multiple rounds of alternating Pauli measurements and conditional Clifford operations---they are not considered in this work.
\par
We define a \emph{stabilizer channel decomposition} of a quantum channel $\bs{\chi}$ as an expression of the form 
\begin{equation}
\label{eq:1}
\bs{\chi}=\sum_{i} q_{i}\mathbf{S}_{i}
\end{equation}
where each $\mathbf{S}_{i}$ is a stabilizer channel and each $ q_{i}$ is a real scalar. There are many different ways to decompose a given channel; however, every decomposition of a trace-preserving channel has the property $\sum_{i} q_{i}=1$, which is a consequence of the fact that each constituent channel $\mathbf{S}_{i}$ is also trace-preserving. Our representation of arbitrary quantum channels as real linear combinations of stabilizer channels (Clifford plus measurements) should be contrasted with the complex linear combinations of Pauli channels used by Aaronson and Gottesman to represent unitary channels~\cite{Aaronson2004}. Notably, our stabilizer channel decomposition can represent both unitary and non-unitary channels.
\par
Each term in a stabilizer channel decomposition individually describes a physical process. Only when all the coefficients are non-negative can they be interpreted as probabilities of a stochastic physical processes. More generally, the coefficients of the stabilizer decomposition in Eq.~(\ref{eq:1}) may be viewed as a quasiprobability distribution, i.e., a distribution in which some of the `probabilities' are negative. The use of quasiprobability distributions to describe quantum mechanical phenomena traces back to the development of the Wigner function~\cite{wigner1932quantum}, wherein the appearance of a negative value indicates non-classical behavior~\cite{kenfack2004negativity}. More recently, quasiprobabilities have been used to express quantum gates and circuits with the finding that negative coefficients signify quantum behavior and strongly affect the complexity of simulating such circuits~\cite{Hofmann2009, Pashayan2015}. Accordingly, an important metric of a stabilizer decomposition is the total magnitude of its negative coefficients, or \emph{negativity}:
\begin{equation}
\label{eq:eta}
\eta \equiv \sum_{i:\  q_i<0} \left|  q_i \right|.
\end{equation}
For a trace-preserving channel the negativity of a decomposition is directly related to its 1-norm:  $\sum_{i} \left| q_i \right| = 1 + 2 \eta \ge 1$. As will be elaborated in Section \ref{sec:perf}, the time complexity of our simulation method is strongly dependent on the 1-norms (and hence the negativities) of the decompositions of the circuit operations.
\par
As a demonstration of the stabilizer channel decomposition, we present two examples of quantum channels that do not admit exact representations as probabilistic mixtures of stabilizer operations but may be represented exactly by (quasiprobability) stabilizer channel decompositions. 
In these examples, $\mathbf{I}$ denotes the identity channel, $\mathbf{Z}$ denotes the channel corresponding to conjugation by the Pauli operator $Z$, and $\mathbf{S}$ denotes the channel corresponding to conjugation by the phase gate $S = \sqrt{Z}$. 
\par
\begin{example}
The channel $\mathbf{Z}_\theta$ corresponding to the coherent rotation $\rho\rightarrow Z_\theta\rho Z_\theta^{\dagger}$,
where $Z_\theta=\left[\begin{array}{cc}
1 & 0\\
0 & e^{i\theta}
\end{array}\right]$, can be decomposed as
\begin{equation}
 \mathbf{Z}_\theta  =\frac{1+\cos\theta-\sin\theta}{2}\mathbf{I}+\frac{1-\cos\theta-\sin\theta}{2}\mathbf{Z}+\sin\theta\mathbf{S}.
\label{eqn:Z decomp2}
\end{equation}
For $0\le\theta\le\pi/4$, this decomposition has the least negativity of any exact decomposition. A prominent special case is $\theta=\pi/4$, which corresponds to the $T$ gate:
\begin{equation}\label{eq:T gate decomp}
\mathbf{T} = \frac{1}{2}\mathbf{I}+\left(\frac{1}{2}-\frac{1}{\sqrt{2}}\right)\mathbf{Z}+\frac{1}{\sqrt{2}}\mathbf{S}.
\end{equation}
\end{example}

\begin{example}
The amplitude damping channel 
\begin{equation}
\label{eq:admodel}
\mathbf{A}_\gamma(\rho) = E_{0}\rho E_{0}^{\dagger}+E_{1}\rho E_{1}^{\dagger}
\end{equation}
with $E_{0}=\left[\begin{array}{cc}
1 & 0\\
0 & \sqrt{1-\gamma}
\end{array}\right]$, $E_{1}=\left[\begin{array}{cc}
0 & \sqrt{\gamma}\\
0 & 0
\end{array}\right]$, and damping parameter $0 \leq \gamma \leq 1$ can be decomposed as
\begin{gather}
\label{eq:addecom}
\mathbf{A}_\gamma = \frac{(1-\gamma)+\sqrt{1-\gamma}}{2}\mathbf{I}+\frac{(1-\gamma)-\sqrt{1-\gamma}}{2}\mathbf{Z}+\gamma\mathbf{R}_{Z}.
\end{gather}
where the coefficient of $\mathbf{Z}$ is negative and approximately $-\gamma/4$ for $\gamma\ll1$.  This decomposition has the least negativity of any exact decomposition.  
\end{example}
\par
As stated, Lemma~\ref{thm:channel-representation} applies to channels involving a fixed number of qubits, but the conclusion also applies to channels acting on any finite-dimensional quantum state space. This is because a channel that maps states in a $d$-dimensional Hilbert space to states in a $d^{\prime}$-dimensional Hilbert space can be embedded in a Hilbert space of $n=\ceil{\log_{2}\max(d,d^{\prime})}$ qubits, to which Lemma~\ref{thm:channel-representation} directly applies. Furthermore, the lemma can be extended to processes with observable outcomes (i.e. measurements). A well-known result of Neumark is that any quantum process with $m$ possible outcomes can be cast as a unitary operation involving the system in question and an auxiliary system with $m$ states, followed by projective measurement on the auxiliary system~\cite{peres1990neumark,rabelo2006algorithm}. It then follows from Lemma~\ref{thm:channel-representation} that such a process can be expressed as a linear combination of Clifford channels acting on the original system plus $\ceil{\log_2 m}$ auxiliary qubits, followed by Pauli measurements of the auxiliary qubits. These conclusions are summarized by the following theorem:
\begin{theorem}
\label{thm:process-representation}
Any quantum process in a $d$-dimensional Hilbert space with $m$ possible outcomes can be expressed exactly as a quasiprobability distribution over stabilizer channels on $\ceil{\log_2 d} + \ceil{\log_2 m}$ qubits, where the stabilizer channels consist of (1) Clifford operations and (2) Pauli measurements followed by Clifford operations conditioned on measurement outcomes.
\end{theorem}
\par
For any number of qubits $n$, the number of Clifford channels substantially outnumbers the dimension of the channel space. Thus there are many different ways to decompose a given channel. Which way is best will depend on the application. Decompositions with the fewest terms may be optimal for analyses that involve a complete enumeration of stabilizer terms. For our simulation method outlined below, decompositions with the smallest 1-norm are optimal as they minimize the time cost. Finding the stabilizer channel decomposition of an arbitrary quantum channel $\bs{\chi}$  that minimizes the 1-norm can be posed as the linear programming problem
\begin{subequations}
\label{eq:lp}
\begin{align}
\min_ q \sum_i | q_i| \\
\intertext{subject to}
\bs{\chi} = \sum_i  q_i \bs{S}_i
\end{align}
\end{subequations}
where $i$ indexes the Pauli reset and Clifford stabilizer channels, $ q_i \in \mathbb{R}$, and $\bs{S}_i$ is the $i$-th stabilizer channel.
\par
For the case of $n = 1$, the linear programming problem in Eq.~(\ref{eq:lp}) involves a system of 16 equations in 30 variables (one for each of 24 Cliffords and one for each of 6 Pauli resets).
For $n = 2$ there are 256 equations in 11550 variables (11520 Cliffords and 30 Pauli resets), where each equation is sparse with exactly 16 nonzeros when channels are expressed as Pauli transfer matrices \cite{Chow2012}. Many relevant quantum logic gates as well as physical noise models may be expressed in terms of 1- or 2-qubit interactions. The 2-qubit channel models may be composed together to generate models for larger channels as needed. However, for $n>2$, the system of equations quickly becomes impractically large with almost 93 million variables for $n=3$. This same difficulty arises for approaches that use a positive decomposition to represent the channel.
\section{Monte Carlo Simulation of Stabilizer Channel Decompositions}
\label{sec:mc}
We now consider the problem of calculating the expectation value for an observable $\phi$ on the final state of a quantum circuit. Generalization to the case of multiple observables is straightforward. We model an initial state $\rho$ acted upon by a sequence of quantum channels $\bs{\chi}^{(1)},\ldots,\bs{\chi}^{(K)}$ that express the circuit gates and error processes. The purpose of the simulation is to compute the expectation value
\begin{gather}
F = \vec{\phi}^\dagger \bs{\chi}^{(K)}\cdots\bs{\chi}^{(1)} \vec{\rho}.
\end{gather}
We assume that each channel $\bs{\chi}^{(k)}$ has been decomposed into a weighted sum of stabilizer channels as
$\bs{\chi}^{(k)}=\sum_{i} q_{i}^{(k)}\mathbf{S}_{i}$, cf.~Eq.~(\ref{eq:lp}).
Similarly, we assume that the initial state $\rho$ and the observable $\phi$ have been decomposed into weighted sums of stabilizer states ${\sigma_i}$ \cite{garcia2015simulation,Yoder2012}:
$\rho=\sum_{i} q_{i}^{(0)} \sigma_{i}$ and $\phi = \sum_{i} q_{i}^{(K+1)}\sigma_{i}$.
\par
With respect to these stabilizer decompositions, the expectation value $F$ may be expressed as
\begin{align} \label{eq:Fsum}
F &= \sum_{i_{K+1}} \cdots \sum_{i_{0}}  q_{i_{K+1}}^{(K+1)} \cdots  q_{i_{0}}^{(0)} \left(  \vec{\sigma}_{i_{K+1}}^\dagger \mathbf{S}_{i_{K}} \cdots \mathbf{S}_{i_{1}} \vec{\sigma}_{i_{0}}\right) \\
 &= \sum_{i}  q(i) f(i)
\end{align}
where $i=(i_{0},\ldots,i_{K+1})$, $ q(i) =  q_{i_{K+1}}^{(K+1)}\cdots q_{i_{0}}^{(0)}$, and $f(i) = \vec{\sigma}_{i_{K+1}}^\dagger \mathbf{S}_{i_{K}} \cdots \mathbf{S}_{i_{1}} \vec{\sigma}_{i_{0}}$. The expression for $f(i)$ describes a stabilizer circuit, hence each $f(i)$ can be computed efficiently.  However, explicit computation of Eq.~(\ref{eq:Fsum}) is impractical as the number of terms is generally exponential in $K$. Instead, $F$ is estimated by sampling terms from the sum. Let $p$ be any probability distribution over $i$ such that $p(i) > 0$ when $q(i) > 0$, and let $i^{(1)},\ldots,i^{(N)}$ denote values of $i$ drawn independently from $p$. Then
\begin{gather}
\tilde{F}_{N} = \frac{1}{N} \sum_{s=1}^{N} \frac{ q(i^{(s)}) f(i^{(s)})}{p(i^{(s)})} \label{f_tilde}
\end{gather}
is an unbiased estimate of $F$, i.e.\ $\langle \tilde{F}_{N} \rangle = F$.  Provided $p$ can be sampled efficiently, $\tilde{F}_{N}$ can be computed efficiently. For reasons discussed in the next section,  we take  $ p(i) =  p^{(K+1)}_{i_{K+1}} \cdots p^{(0)}_{i_0} $ where $p^{(k)}_i \equiv | q^{(k)}_i| / \sum_j | q^{(k)}_j|$. The complete simulation procedure is summarized in Algorithm \ref{alg:simulation algorithm}.

\begin{algorithm}
\caption{Observable estimation in terms of stabilizer states and stabilizer channels.}
\label{alg:simulation algorithm}
\begin{algorithmic}
\Require An initial state $\rho$ with stabilizer state decomposition $\rho=\sum_{i} q_{i}^{(0)} \sigma_{i}$
\Require A sequence of channels $\bs{\chi}^{(1)}, \ldots, \bs{\chi}^{(K)}$  with stabilizer channel decompositions $\bs{\chi}^{(k)}=\sum_{i} q_{i}^{(k)}\mathbf{S}_{i}$
\Require An observable $\phi$ with stabilizer state decomposition $\phi=\sum_{i} q_{i}^{(K+1)} \sigma_{i}$
\Require The number $N$ of runs to simulate
\State Let $p^{(k)}_i \equiv | q^{(k)}_i|/ \sum_j | q^{(k)}_j|$ for $k=0,1,\ldots,K+1$.
\State  $\tilde{F} \leftarrow 0$
\For {$r = 1$ to $N$}
	\For {$k=0$ to $K+1$}
		\State Select $i_k$ with probability $p^{(k)}_i$
	\EndFor
	\State $\rho_{*} \leftarrow \sigma_{i_0}$
	\For {$k = 1$ to $K$}
		\State $\rho_{*} \leftarrow \bs{S}_i(\rho_{*})$
	\EndFor
	\State $w \leftarrow  q^{(0)}_{i_0} \cdots  q^{(K+1)}_{i_{K+1}} / p^{(0)}_{i_0} \cdots p^{(K+1)}_{i_{K+1}}$
	\State $\tilde{F} \leftarrow \tilde{F} + w \vec{\sigma}_{i_{K+1}}^\dagger \vec{\rho}_{*} / N$
\EndFor
\Ensure $\tilde{F} \approx \vec{\phi}^\dagger \bs{\chi}^{(K)} \cdots \bs{\chi}^{(1)} \vec{\rho}$
\end{algorithmic}
\end{algorithm}
\section{Performance and Scaling} 
\label{sec:perf}
The precision of Monte Carlo simulation can be quantified by the variance of the estimator $\tilde{F}_N$:
\begin{align}
\operatorname{Var}\tilde{F}_{N} &\equiv \left\langle |F^2| \right\rangle - {\langle F \rangle}^2 \\
&= \frac{1}{N} \left(\left< \left| q f/p\right|^{2} \right> - \left|F\right|^{2}\right) = \frac{1}{N} \operatorname{Var}\tilde{F}_{1}.
\end{align}
(Here the random variable underlying the expectation values is the argument $i$ common to the functions $p$, $q$, and $f$.) The number of samples $N$ needed to achieve a specified variance scales as $\operatorname{Var}\tilde{F}_{1}$, which
is determined by the nature of the circuit being simulated as well as the choice of the sampling distribution $p$.
The minimal value of $\operatorname{Var}\tilde{F}_{1}$ is obtained by sampling from the distribution $p^{*}(i)=\left| q(i) f(i)\right| / \sum_{j}\left|  q(j) f(j)\right|$.
However, computing $p^{*}$ is typically as difficult as computing $F$ itself.
Instead, we use a distribution that holds to the key principle that $p$ should be large (small) when $| q f|$ is large (small). 
Our choice for $p$ described above is the product of the optimal sampling distributions for the individual channels, which simultaneously satisfies this key principle and remains easy to sample from.
\par
Let $g_{k} \equiv \sum_i q^{(k)}_i$ denote the 1-norm of the $k$th stabilizer decomposition ($k=0,1,\ldots,K+1$). Using the fact that
\begin{gather}
\frac{  q(i) }{ p(i) } = \operatorname{sign} \left(  q_{i_{K+1}}^{(K+1)} \cdots  q_{i_{0}}^{(0)}\right) g_{K+1} \cdots g_{0}
\end{gather}
the variance can be expressed as
\begin{align}
\label{eq:var}
\operatorname{Var}\tilde{F}_{N} &=\frac{1}{N}\left( g_{K+1}^2 \cdots g_{0}^2 \left< \left|f \right|^{2} \right> - \left| F\right|^{2}\right).
\end{align}
Now, suppose $f(i)$ were replaced everywhere by $f(i)-\frac{1}{2}$. This would shift the expectation value of $\tilde{F}_N$ by a constant $\Delta$, but would not alter its variance. Thus
\begin{align}
\operatorname{Var}\tilde{F}_N &= \frac{1}{N}\left( g_{K+1}^2 \cdots g_{0}^2  \left< \left|f - \textstyle\frac{1}{2} \right|^{2} \right> - \left| F - \Delta \right|^{2}\right) \\
 &\le \frac{ g_{K+1}^2 \cdots g_{0}^2 }{N}   \left< \left|f - \textstyle\frac{1}{2} \right|^{2} \right>.
\end{align}
Recall that $f(i)$ is the projection of one stabilizer state onto another. Since such a projection lies in the interval $[0,1]$, we have $|f(i) - \frac{1}{2}| \le \frac{1}{2}$ for all $i$. It follows that
\begin{align}
\operatorname{Var}\tilde{F}_N \le \frac{ g_{K+1}^2 \cdots g_{0}^2 }{4N}
\end{align}
which is in fact a tight bound. To make the scaling of $N$ with circuit size more explicit, we define $g_{\text{in}} \equiv g_{0}$, $g_{\text{obs}} \equiv g_{K+1}$, and $ g_{\text{ch}} \equiv \max_{k=1,\cdots,K} g_k \ge 1$. Then the number of simulations $N$ needed to estimate $F$ to a given statistical uncertainty $\epsilon$ is
\begin{gather} \label{eq:N bound from variance}
N \le\frac{  g_{\text{in}}^2 g_{\text{obs}}^2 g_{\text{ch}}^{2K}  }{4\epsilon^2} .
\end{gather}
A similar bound can be obtained using Hoeffding's inequality, which for the present problem is
\begin{gather}
\Pr(|\tilde{F}_{N}-F|>\epsilon) \le \exp\left(-\frac{2N\epsilon^2}{  g_{\text{in}}^2 g_{\text{obs}}^2 g_{\text{ch}}^{2K} } \right)
\end{gather} 
Rearranging this expression yields a bound on the number of simulations $N$ needed to ensure that the probability of obtaining an estimate with error  $\le \epsilon$ is at least $1-\delta$:
\begin{gather} \label{eq:N bound from Hoeffding}
N \le \frac{ g_{\text{in}}^2 g_{\text{obs}}^2 g_{\text{ch}}^{2K} }{2\epsilon^2}   \ln \frac{2}{\delta}
\end{gather}
Equations ~(\ref{eq:N bound from variance}) and (\ref{eq:N bound from Hoeffding}) indicate that simulation cost as measured by the number of runs required is generally exponential in the circuit size $K$.
However the base $g_{\text{ch}}$ is often close to 1. 
In particular, if each channel happens to be exactly representable as a positive sum of stabilizer channels, then $g_{\text{ch}} = 1$ and the variance does not increase with $K$.
In that case our simulation method reduces to existing methods that simulate circuits as probabilistic mixtures of stabilizer circuits.
For comparison, the 1-norm for optimal decomposition of the $T$ gate given in Eq.~(\ref{eq:T gate decomp}) is $\sqrt{2}$. Each $T$ gate doubles the variance of the estimator and hence doubles the cost of obtaining an estimate of given precision.
The reason the variance grows exponentially when the decompositions involve negative coefficients is that the expected value is obtained as an interference between much larger terms of opposite sign.
Any imbalance between the number of sampled positive terms versus the number of sampled negative terms results in a relatively large error in the estimated value.
This situation may be seen as a version of the well-known ``sign problem'' encountered in calculations of the properties of fermionic systems \cite{Troyer2005}.
\par
The targeted application is simulation of a circuit of stabilizer gates with weak (i.e., rare and/or close to identity) non-stabilizer errors. For the error channels considered in the examples, the negativity of the stabilizer decompositions are within small factors of commonly used error measures, such as infidelity and trace distance. We thus have $g_{\text{ch}}^2 \sim 1 + p_{\text{NC}}$ where $p_{\text{NC}} \ll 1$ may be loosely interpreted as  the per-gate probability of a non-Clifford error. The number of runs needed to obtain a fixed accuracy estimate scales as $g_{\text{ch}}^{2K} \sim e^{K p_{\text{NC}}}$, which does not increase rapidly until $K  p_{\text{NC}} \gtrsim 1$. Thus as a rule-of-thumb, our simulation method is efficient for $K p_{\text{NC}} \lesssim 1$. For example, if the per-gate probability of a non-Clifford error is $\sim10^{-3}$, using our method one can accurately and efficiently simulate circuits of at least $\sim1000$ gates. In contrast, simulations using positive-mixture approximations can be guaranteed accurate only for $K p_{\text{NC}} \ll 1$. Our method therefore makes it feasible to accurately simulate quantum circuits for which efficient stochastic methods may be significantly inaccurate. Additionally, our method supports a continuous trade-off between cost and model accuracy through the choice of how much negativity is used to model non-Clifford channels.
\par
We have so far discussed the simulation cost in terms of the number of runs needed to obtain an estimate of specified precision The time cost of each run is just that of stabilizer propagation: proportional to the number of gates, and between linear and quadratic in the number of qubits \cite{Aaronson2004}. There is additionally the cost of decomposing the initial state, the channels representing the noisy gates of the circuit, and the final observable. If there is a fixed 1- or 2-qubit error model for each type of gate and there are only a few types of gates, these decompositions can be performed once up front with negligible amortized cost. Finally, we observe that if there are multiple observables to be computed, only the last step of each circuit simulation needs to be repeated for each observable. Thus the total time cost for simulating a circuit consisting of $n$ qubits, $K$ quantum channels, and $M$ observables scales as $O(n^2 (K + M) (1+p_{\text{NC}})^{K})$. The space cost is also essentially that of stabilizer propagation, $O(n^2)$ \cite{gottesman1998heisenberg}. In practice this simulation method is time-limited rather than memory-limited.
As with other Monte Carlo methods, it is embarrassingly parallel and therefore the time cost can be greatly reduced with parallel computation.
\section{Implementation} 
\label{sec:ex}
We demonstrate the accuracy of this stabilizer decomposition using a numerical implementation of Algorithm~\ref{alg:simulation algorithm}. Our implementation relies on a heavily modified version of the CHP stabilizer circuit simulator \cite{Aaronson2004}, which propagates quantum states represented as a tableau of signed Pauli operators. Our modifications offer support for non-unital channels by implementing the Pauli $Z$ reset gate as well as methods to efficiently compute the projection of a stabilizer state onto any given stabilizer subspace.
As an added benefit, support for measurement of multi-qubit Pauli operators has a time cost $O(n^2)$ with the resulting projection function scaling as $O(n^3)$. This is comparable with the state of the art methods from Cross et al.~\cite{Cross2014} but with the advantage that an intermediate circuit does not need to be synthesized to reduce the tableau to a canonical form.
\par
The modified stabilizer propagation kernel is embedded in our simulator software responsible for implementing circuit scheduling, fault injection, and error correction logic. The CHP simulator is repeatedly executed to collect the statistics needed to estimate system observables by Monte Carlo sampling. Within the simulator, a parser decomposes quantum circuits into a discrete schedule of gates acting on addressable qubits. We use the well-known QASM pseudo-code notation to specify the input circuit and the internal representations of the simulator \cite{Balensiefer2005}. A gate scheduler modifies this circuit description to add subcircuit boundaries, i.e., timestamps, to identify the order in which gates act on qubits. Along each boundary, we insert noise operators that model the errors caused by each subcircuit. The stabilizer decomposition for each gate and noise operator is generated and stored by the simulator.
\par
When an instance of a circuit is simulated, each gate is represented by an element from its stabilizer decomposition. These elements are chosen randomly with respect to the probability distribution determined by the stabilizer decomposition coefficients. An instance of a stabilizer circuit is used as a Monte Carlo sample for the non-stabilizer channel. We translate each stabilizer element in the circuit instance into a sequence of C, H, P, measure, and reset operations. The resulting sequence of CHP operations is simulated using the stabilizer propagation kernel. In the simulations of the Steane code described below, the simulator frequently encounters gates conditioned upon previous measurement outcomes, i.e., syndromes. In our implementation, syndromes are returned by the propagation kernel to the circuit simulator and decoded using a pre-computed lookup table. The gate scheduler prepares the subcircuit that implements the corresponding error correction operation and pushes this new gate sequence to the propagation kernel.
\par
For the numerical demonstrations reported here, we focus on estimating the infidelity of a qubit logically encoded using a quantum-error correction code and stored in the presence of physical noise. Therefore, we conclude each simulation instance by computing the projection of the final state onto a specified (error-free) code state. For calculating this fidelity, we project the simulated noisy state onto the ideal state prepared by the noiseless circuit. We accumulate a running average of the fidelity with the $i$-th result weighted by the factor $ q_{i}$, as determined by the coefficients of the $i$-th randomly selected noisy, circuit instance. A complete simulation repeats the process of sampling the stabilizer channels, propagating the circuit instance, and calculating the desired observable $N$ times, where the number of samples $N$ is specified as input.
\par
\begin{figure}
\centering
\includegraphics[width=\columnwidth]{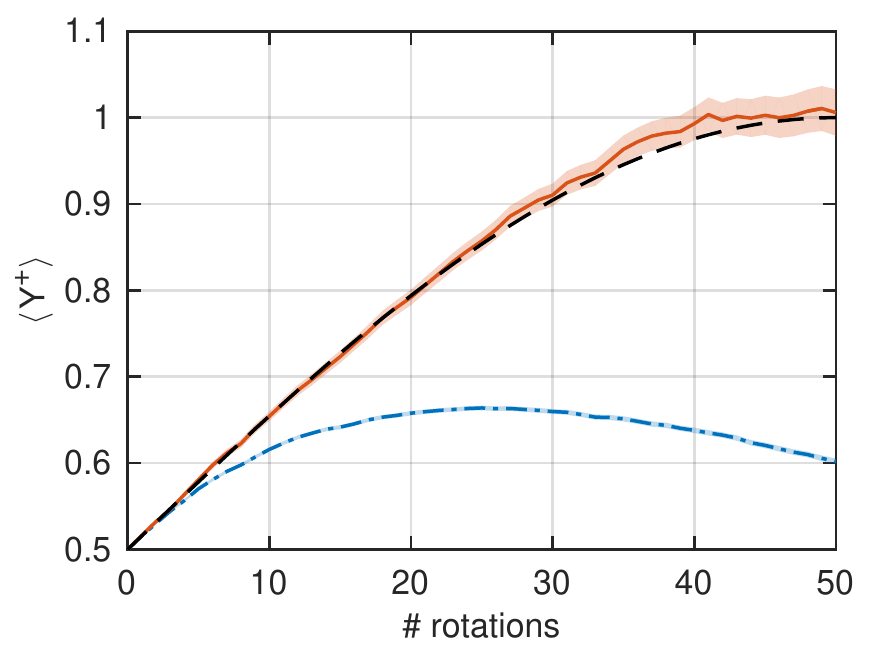}
\caption{Monte Carlo simulation of a qubit undergoing repeated coherent rotations, as revealed by its projection onto the +1 eigenstate of the Pauli $Y$ operator.  (dashed black line) The analytical result $\langle Y^{+} \rangle = (1+\sin\theta)/2$. (solid red line) Estimate using an exact stabilizer decomposition of the rotation channel. (dash-dot blue line) Estimate using an approximate positive decomposition. Shaded bands cover the area one standard deviation above and below the estimates.}
\label{fig: qubit rotation}
\end{figure}
We have verified the correctness of our numerical simulator by comparing its results for several test circuits with exact results obtained from an independent density matrix simulator. The results from Monte Carlo simulation were observed to converge to the results of the corresponding density matrix simulation for all test cases provided sufficient samples were collected. As a proof-of-principle example that also illustrates the advantages of this approach, we simulated a simple circuit in which a qubit initially in the $+1$ eigenstate of the Pauli $X$ operator is rotated in 50 equal steps to the $+1$ eigenstate of the Pauli $Y$ operator. Each rotation operator $Z_{\pi/100}$ is a non-stabilizer operation with an exact decomposition given by Eq.~(\ref{eqn:Z decomp2}). The expected value of $Y^{+}$ was computed at each time step (i.e.~after each rotation) and added to a running average, yielding a Monte Carlo estimate of the time dependence of  $\langle Y^{+}\rangle$. A typical outcome for 10,000 Monte Carlo runs is shown in Fig.~\ref{fig: qubit rotation}). In this case the final value of $\langle Y^{+} \rangle$ was estimated as $1.006\pm0.027$, which is in statistical agreement with correct value (1). While it might seem conspicuous that the estimated probability exceeds 1, the fact that the estimator is unbiased means that if the estimate has any chance of being less than the mean it must also have a chance of exceeding the mean.
\par
For comparison, we also simulated the circuit using the approximate decomposition $\mathbf{Z}_{\pi/100} \approx (1-\sin\theta) \mathbf{I} +\sin\theta \mathbf{S}$, which is the positive decomposition that best reproduces the initial evolution of $\langle Y^{+} \rangle$. In a typical run of 10,000 samples we obtained an estimate with mean $0.602$ and standard deviation $0.003$  (Fig.~\ref{fig: qubit rotation}. Clearly, the use of an approximate positive decomposition in this case yields a strongly biased estimator. In addition, the estimator's precision is a very misleading indicator of its accuracy. We further observed that the estimated state using the exact channel decomposition remains approximately pure, whereas the estimated state using the approximate positive decomposition becomes more mixed with each step.


\section{Simulations of Steane [[7,1,3]] Circuits} 
\label{sec: Steane results}

\begin{figure*}[t]
\centering
\includegraphics[width=6in]{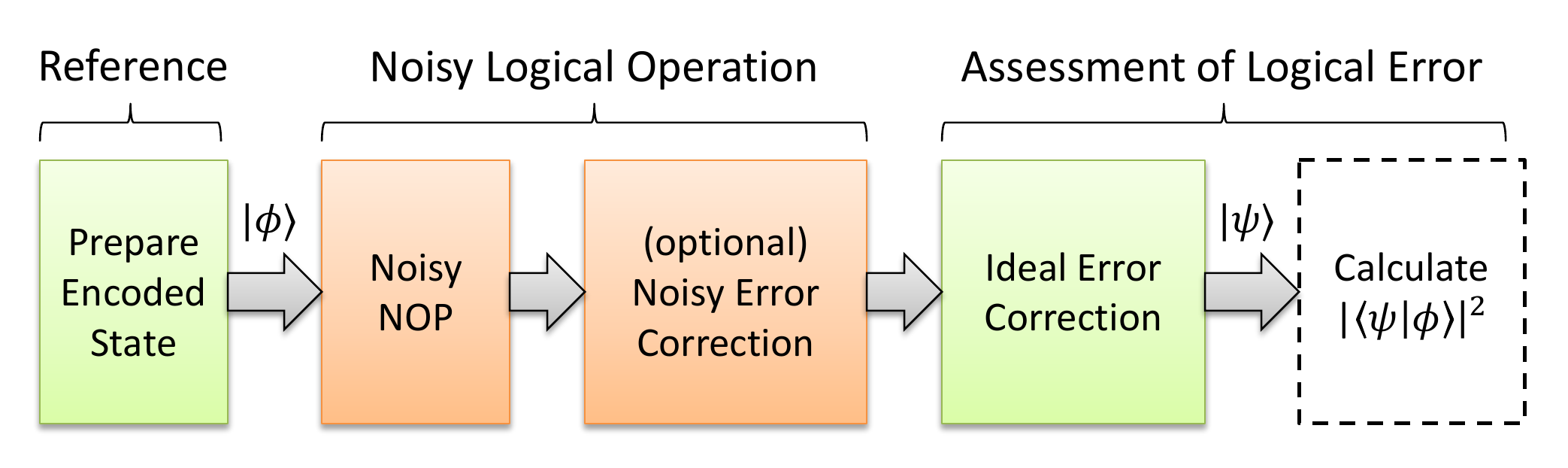}
\caption{Block structure of the circuit simulated as demonstration of our simulation method. The circuit prepares a reference logical state in the Steane [[7,1,3]] encoding. The encoded state is then subjected to errors, and optionally, a round of noisy error correction. The logical error of the resulting state is obtained by removing correctable errors with a round of error-free error correction, then projecting the result onto the originally encoded state. Details of the subcircuits, including specific gates and locations of error insertion, are provided in Appendix B.}
\label{fig:circuit overview}
\end{figure*}

We used the numerical implementation of Algorithm 1 to investigate the logical error rate for a single-level encoding of the Steane [[7,1,3]] quantum error correction code in the presence of either depolarizing or amplitude damping noise. We used the former noise model as a baseline test case since the depolarizing noise may be represented as a non-negative mixture of stabilizer operators. It is possible to simulate depolarizing noise using conventional stabilizer propagation and we used this as a check for our numerical implementation. We used the latter case of amplitude damping as a test of the applicability of our new method for simulating weakly non-Clifford noise model as given in Eq.~(\ref{eq:addecom}). To confirm the correct implementation of these noise models in our algorithm, we simulated each noise channel on various 1-qubit input states and found that results matched analytical predictions.
\par
The input to the Steane circuit simulations was a single-qubit state to be prepared as a logical qubit using a noiseless encoding circuit. The encoding was followed by a noisy logical identity operation during which noise may occur on any qubit. Error correction was then performed consisting of syndrome measurement, decoding, and applying the corresponding correction operation. These correction operations were assumed to be noisy as well. To calculate the logical fidelity of this operation, an additional round of noiseless error correction was performed and the fidelity of the resulting output state was taken with respect to the noiseless, logically encoded input. The specific encoding, syndrome measurement, and decoding circuits used in this numerical study are presented in Appendix \ref{app:circuits}.
\par
We calculated the logical error of the Steane encoded state as the probability that a noisy logical operation followed by noisy QEC operations would yield an unrecoverable error in the logical state. Therefore, we did not include noise within the encoding circuit in order to ensure the input logical state is error-free. In addition, following the noisy error correction operation in Fig.~\ref{fig:circuit overview}, we performed a second round of noiseless error correction. The purpose of the second round of QEC was to remove any correctable errors in the logical state. The logical error was then computed as the infidelity between the two logical states. The infidelity was calculated for six different input states, specifically the eigenstates of the Pauli operators $X,Y,Z$, and averaged. In the discussion of our results, ``infidelity'' always refers to the output state infidelity averaged over the six Pauli input states. 

\begin{figure}
\centering
\includegraphics[width=\columnwidth]{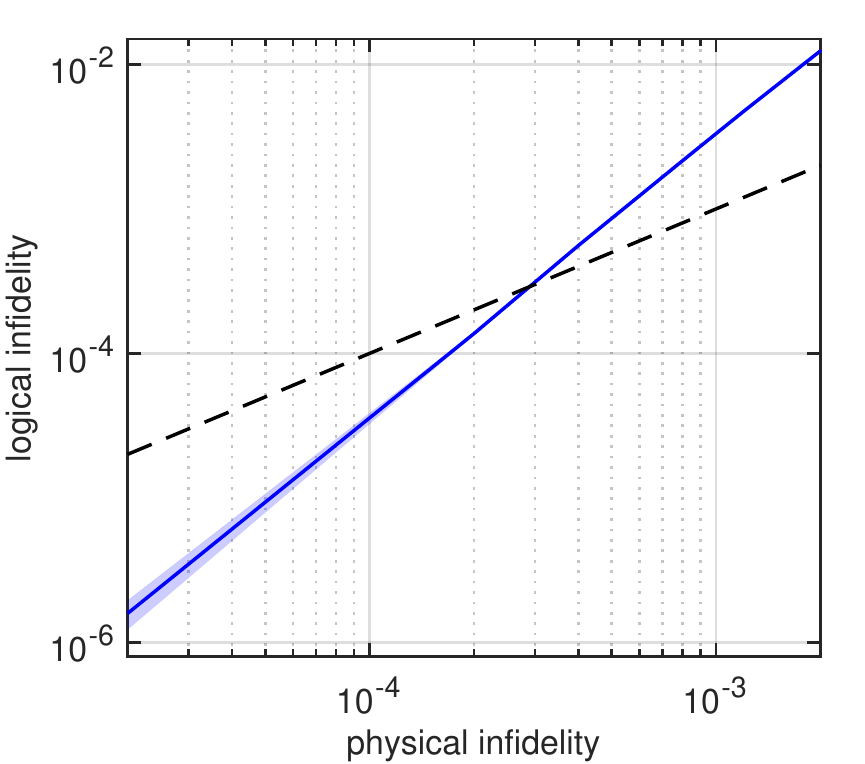}
\caption{(sold blue line with shaded uncertainty band) Estimated infidelity of a  Steane [[7,1,3]] logical identity operation as a function of physical qubit depolarization.  The circuit implementing the logical operation is specified by Fig.~\ref{fig:circuit overview}. For comparison, the physical infidelity is shown by the dashed black line.}
\label{fig:dpler}
\end{figure}
\par
Our first example considers the depolarizing noise model, in which the channel
\begin{equation}
\mathbf{D}(\rho) = (1-p) \mathbf{I} + \frac{p}{3}\left(\mathbf{X} + \mathbf{Y} + \mathbf{Z}\right).
\end{equation}
is applied to each qubit involved in a noisy operation. The infidelity $\bar{I}= 1-\bar{F}$ of a qubit under this channel is $2p/3$. Fig.~\ref{fig:dpler} shows the estimated infidelity of the logical (Steane-encoded) identity operation as a function of the physical (1-qubit) infidelity. The solid line is the estimate $\tilde{\Phi}$; the shaded band marks the region $\tilde{\Phi} \pm \epsilon$ where $\epsilon$ is the standard deviation of $\tilde{\Phi}$ estimated from the sample variance. Our simulations indicate that the logical infidelity is less than the physical infidelity (shown by a dashed black line) when the physical infidelity is less than approximately $2.7\times 10^{-4}$ ($p\lesssim 4\times 10^{-4}$). This result agrees very well with the average level-1 pseudo-threshold computed in \cite{Gutierrez2015} (row ``DC'' of Table VII therein).
\par
Our second example considers the amplitude damping noise model, wherein the channel $\mathbf{A}$ defined in eq.~(\ref{eq:admodel}) is applied to each qubit involved in a noisy operation. For our simulations we used the stabilizer decomposition given by Eq.~(\ref{eq:addecom}), whose negativity is approximately $-\gamma/4$, where $\gamma$ ranges from $0$ to $1$.   For $\gamma\ll 1$ the infidelity of this channel is $\bar{I}\approx \gamma/3$, which was obtained the relation $\bar{F} = (2 F_\mathbf{A,I} + 1)/3$ \cite{Nielsen2002}, where $F_\mathbf{A,I} = (1+\sqrt(1-\gamma))^2/4$ is the fidelity of the process matrices for channels $\mathbf{A}$ and $\mathbf{I}$.  Fig.~\ref{fig:dpler} shows the estimated logical infidelity as a function of the physical infidelity under the amplitude damping noise model. Again, the solid line denotes the estimate, the shaded band shows the estimated uncertainty, and the dashed line shows the physical infidelity. The slight bend in the logical infidelity appears to be an artifact of statistical fluctuations. The logical infidelity is less than the physical infidelity when the latter is less than approximately $2.9 \times 10^{-4}$ ($\gamma \lesssim 9 \times 10^{-4}$).  This is about a factor of two greater than the pseudo-threshold estimated by Gutierrez et al.~for a similar circuit \cite{Gutierrez2015}. This difference may be attributed to several differences in our approach: Whereas we use majority vote when three rounds of syndrome extraction are performed, Gutierrez et al.~rely on whatever results arises from the third round. In addition, they use the trace distance as the basis for threshold calculations and average over a larger set of input states. 
\begin{figure}
\centering
\includegraphics[width=\columnwidth]{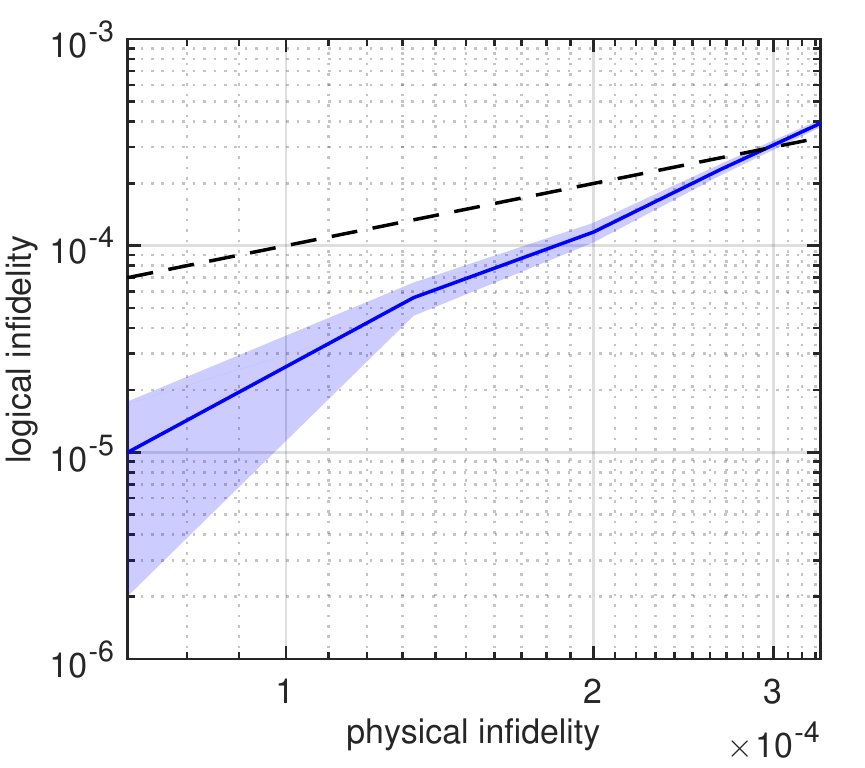}
\caption{(sold blue line with shaded uncertainty band) Estimated infidelity of a  Steane [[7,1,3]] logical identity operation as a function of physical qubit amplitude damping.  The circuit implementing the logical operation is specified by Fig.~\ref{fig:circuit overview}. For comparison, the physical infidelity is shown by the dashed black line.}
\label{fig:adhler}
\end{figure}
\section{Conclusion} 
\label{sec:conc}
Modeling and simulation play an important role in verifying and validating quantum information technologies. We have presented a new method for simulating arbitrary quantum circuits that leverages the efficiency of stabilizer propagation and Monte Carlo sampling. A key aspect of this method is that it offers an exact representation of the circuit described by quasiprobability distributions over efficiently simulatable operations. The cost of this simulation method depends smoothly on the negativity of the quasiprobability distributions, which may serve as a measure of their non-Cliffordness. Consequently, this simulation method is particularly well suited for studying quantum error correction circuits with weak non-Clifford noise models. A promising application is the study of coherent error models which can be expressed exactly and simulated using our methods. Coherent errors in quantum error correction have recently shown interesting results \cite{Darmawan2016,Suzuki2017,Barnes2017}.
\par
Unbiased Monte Carlo simulation of the stabilizer channel decomposition offers several convenient features over existing simulation methods. This includes a memory scaling that is polynomial in circuit size while maintaining an exact description of the quantum circuit. The Monte Carlo sampling method is highly parallelizable and portable to large-scale distributed computing environments. In addition, the time scaling is controllable via the number of samples requested to achieve a desired estimator variance. Finally, by limiting the amount of negativity allowed in the quasiprobability representations of circuit operations, the trade-off between simulation accuracy and efficiency is adjustable. Quantifying this trade-off is an important direction for future work to better understanding the relationship between the nature of a quantum channel and its optimal quasiprobability representation. Future work will also include an investigation into efficient methods for obtaining optimal or near-optimal stabilizer decompositions of quantum channels involving two or more qubits.
\section*{Acknowledgments}
This worked was sponsored by the Intelligence Advanced Research Project Activity. This manuscript has been authored by UT-Battelle, LLC, under Contract No. DE-AC0500OR22725 with the U.S. Department of Energy. The United States Government retains and the publisher, by accepting the article for publication, acknowledges that the United States Government retains a non-exclusive, paid-up, irrevocable, world-wide license to publish or reproduce the published form of this manuscript, or allow others to do so, for the United States Government purposes. The Department of Energy will provide public access to these results of federally sponsored research in accordance with the DOE Public Access Plan.
\bibliographystyle{apsrev}
\bibliography{ms_manuscript}
\appendix
\section{Clifford Decomposition} 
\label{sec:decomp}
We extend the argument made in Ref.~\cite{Kimmel2014} to prove Theorem~\ref{thm:channel-representation}, which states that any trace-preserving quantum channel can be expressed as a real linear combination of stabilizer channels and, specifically, Clifford channels and Pauli measure-reset channels. Let $\mathcal{P}_{n}$ denote the set of unsigned $n$-qubit Pauli operators and let $\bs{1}^{(P^{\prime},P)}$ denote the (generally unphysical) channel that maps $P \in \mathcal{P}_{n}$ to $P^\prime \in \mathcal{P}_{n}$ and maps every other element of $\mathcal{P}_{n}$ to 0. 
Since $\mathcal{P}_{n}$ is a basis for $n$-qubit states, $\{\bs{1}^{(P^{\prime},P)}\}$ is a basis for the space of $n$-qubit channels. 
We will show that each basis channel can be written as a real linear combination of stabilizer channels, and thus that any quantum channel can be expressed as such a linear combination.
\par
We use the fact that each non-identity operator $P\in\mathcal{P}_{n}$ commutes with exactly half of the elements
of $\mathcal{P}_{n}$ and anticommutes with the remaining elements.
Thus 
\begin{gather}
\sum_{Q\in\mathcal{P}_{n}}\comsign{Q,P}=\begin{cases}
4^{n} & P= I^{\otimes n}\\
0 & \text{else}
\end{cases}
\end{gather}
where $\comsign{Q,P}\equiv1\,(-1)$ in the case that $Q$ commutes
(anticommutes) with $P$.  Furthermore, $\comsign{Q,P}\comsign{Q,P^{\prime}}=\comsign{Q,PP^{\prime}}$
which gives
\begin{gather}
\sum_{Q\in\mathcal{P}_{n}}\comsign{Q,P}\comsign{Q,P^{\prime}}=
\begin{cases}
4^{n} & P= P^{\prime}\\
0 & \text{else}
\end{cases}.
\end{gather}
Consider the linear combination of stabilizer channels 
\begin{gather}
\frac{1}{4^{n}}\sum_{Q\in\mathcal{P}_{n}}\comsign{P,Q}\mathbf{Q}
\end{gather}
where $\mathbf{Q}$ is the channel corresponding to operator $Q$.
We have
\begin{align}
\frac{1}{4^{n}}\sum_{Q\in\mathcal{P}_{n}}\comsign{P,Q}\mathbf{Q} (P^{\prime}) & =\frac{1}{4^{n}}\sum_{Q\in\mathcal{P}_{n}}\comsign{Q,P}QP^{\prime}Q^{\dagger}\nonumber \\
 & =P^{\prime}\frac{1}{4^{n}}\sum_{Q\in\mathcal{P}_{n}}\comsign{Q,P}\comsign{Q,P^{\prime}}\\
 & =\begin{cases}
P^{\prime} & P^{\prime}= P\\
0 & \text{else}
\end{cases}\\
 & =\bs{1}^{(P,P)}.
\end{align}
The expression on the left is thus a stabilizer decomposition of $\bs{1}^{(P,P)}$.
\par
We now obtain a decomposition of $\bs{1}^{(P^{\prime},P)}$ where $P^{\prime}\ne P$
and $P,P^{\prime}\ne I^{\otimes n}$. Let $\mathbf{C}$ be any Clifford
channel that maps $P$ to $P^{\prime}$. Such a channel always exists. Then
\begin{gather}
\mathbf{C}\mathbf{1}^{(P,P)}(P^{\prime\prime})=\left\{ \begin{array}{ll}
P^{\prime} & P^{\prime\prime}=P\\
0 & \text{else}
\end{array}\right\} =\mathbf{1}^{(P^{\prime},P)}.
\end{gather}
To obtain a channel of the form $\bs{1}^{(P,I^{\otimes n})}$
where $P\ne I^{\otimes n}$, let $\mathbf{R}_{P}$ denote a channel
that sets the qubits to a +1 eigenstate of $P$. This may be implemented
as a measurement of $P$ conditionally followed by a Clifford channel
$\mathbf{N}_{P}$ that maps $P$ to $-P$ , where $\mathbf{N}_{P}$
is applied only in the case that measurement yielded the $-1$ eigenvalue.
Using the fact that $(I^{\otimes n} \pm P)/2$ is the projector onto the  $\pm1$ eigenspace of $P$, we have
\begin{align}
\nonumber
\mathbf{R}_{P}(I^{\otimes n}) & = \frac{I^{\otimes n}+P}{2} I^{\otimes n} \frac{I^{\otimes n}+P}{2} + \\ 
& \mathbf{N}_{P} \frac{I^{\otimes n}-P}{2} I^{\otimes n} \frac{I^{\otimes n}-P}{2} \\
 & =\frac{I^{\otimes n}+P}{2}+\mathbf{N}_{P} \frac{I^{\otimes n}-P}{2}\\
 & = I^{\otimes n}+P.
\end{align}
Then 
\begin{gather}
\mathbf{R}_{P}\mathbf{1}^{(I^{\otimes n},I^{\otimes n})}(P^{\prime})=\begin{cases}
I^{\otimes n} + P & P^{\prime}=I^{\otimes n}\\
0 & P^{\prime}\ne I^{\otimes n}
\end{cases}
\end{gather}
yields the decomposition
$\bs{1}^{(P,I^{\otimes n})}=\mathbf{R}_{P}\mathbf{1}^{(I^{\otimes n},I^{\otimes n})}-\mathbf{1}^{(I^{\otimes n},I^{\otimes n})}$.
\par
Summarizing, we have 
\begin{align}
\bs{1}^{(P^{\prime},P)} & =\frac{1}{4^{n}}\sum_{Q\in\mathcal{P}_{n}}\comsign{P,Q}\mathbf{CQ}\label{eq: unital basic decomposition}\\
\bs{1}^{(P^{\prime},I^{\otimes n})} & =\frac{1}{4^{n}}\sum_{Q\in\mathcal{P}_{n}}(\mathbf{R}_{P^{\prime}}-1)\mathbf{Q}\label{eq: nonunital basic decomposition}
\end{align}
where $P,P^{\prime}\in\mathcal{P}_{n}$ with $P\ne I^{\otimes n}$,
$\mathbf{C}$ is any Clifford channel that maps $P$ to $P^{\prime}$,
and $\mathbf{R}_{P}$ is any channel that resets $P$ to a +1 eigenvalue.
Each of the expressions above is clearly a linear combination of stabilizer
channels, where each term is a product of at most one Pauli measurement
and at most two Clifford operations. All that remains are channels
of the form $\bs{1}^{(I^{\otimes n},P)}$, and we do not need
to decompose these since a trace-preserving channel cannot have such a channel as a component.
(If it did, the trace-1 states $(I^{\otimes n} + P)/2$ and $(I^{\otimes n} - P)/2$ would map to states of unequal trace.)
Thus the space of trace-preserving $n$-qubit channels is spanned by the set of $n$-qubit
stabilizer channels.
\section{Quantum Circuits for Numerical Simulation} 
\label{app:circuits}
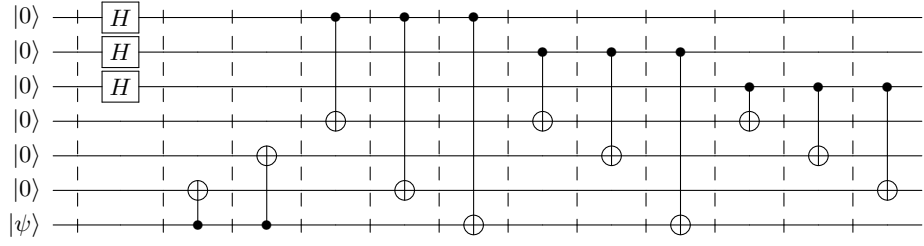
\begin{figure*}[h]
\mbox{
\centering
\Qcircuit @C=1em @R=.15em @!R {
\lstick{\ket{0}}  & \qw\ar@{--}[]+<0em,.3em>;[dddddd]+<0em,-.3em> & \gate{H} & \qw\ar@{--}[]+<0em,.3em>;[dddddd]+<0em,-.3em> & \qw & \qw\ar@{--}[]+<0em,.3em>;[dddddd]+<0em,-.3em> & \qw & \qw\ar@{--}[]+<0em,.3em>;[dddddd]+<0em,-.3em> & \ctrl{3} & \qw\ar@{--}[]+<0em,.3em>;[dddddd]+<0em,-.3em> & \ctrl{5} & \qw\ar@{--}[]+<0em,.3em>;[dddddd]+<0em,-.3em> & \ctrl{6} & \qw\ar@{--}[]+<0em,.3em>;[dddddd]+<0em,-.3em> & \qw & \qw\ar@{--}[]+<0em,.3em>;[dddddd]+<0em,-.3em> & \qw & \qw\ar@{--}[]+<0em,.3em>;[dddddd]+<0em,-.3em> & \qw & \qw\ar@{--}[]+<0em,.3em>;[dddddd]+<0em,-.3em> & \qw & \qw\ar@{--}[]+<0em,.3em>;[dddddd]+<0em,-.3em> & \qw & \qw\ar@{--}[]+<0em,.3em>;[dddddd]+<0em,-.3em> & \qw & \qw\\
\lstick{\ket{0}}  & \qw & \gate{H} & \qw & \qw & \qw & \qw & \qw & \qw & \qw & \qw & \qw & \qw & \qw & \ctrl{2} & \qw & \ctrl{3} & \qw & \ctrl{5} & \qw & \qw & \qw & \qw & \qw & \qw & \qw\\
\lstick{\ket{0}}  & \qw & \gate{H} & \qw & \qw & \qw & \qw & \qw & \qw & \qw & \qw & \qw & \qw & \qw & \qw & \qw & \qw & \qw & \qw & \qw & \ctrl{1} & \qw & \ctrl{2} & \qw & \ctrl{3} & \qw\\
\lstick{\ket{0}}& \qw & \qw & \qw & \qw & \qw & \qw & \qw & \targ & \qw & \qw & \qw & \qw & \qw & \targ & \qw & \qw & \qw & \qw & \qw & \targ & \qw & \qw & \qw & \qw & \qw\\
\lstick{\ket{0}} & \qw &  \qw & \qw & \qw & \qw & \targ & \qw & \qw & \qw & \qw & \qw & \qw & \qw & \qw & \qw & \targ & \qw & \qw & \qw & \qw & \qw & \targ & \qw & \qw & \qw\\
\lstick{\ket{0}} & \qw &  \qw & \qw & \targ & \qw & \qw & \qw & \qw & \qw & \targ & \qw & \qw & \qw & \qw & \qw & \qw & \qw & \qw & \qw & \qw & \qw & \qw & \qw & \targ & \qw\\
\lstick{\ket{\psi}} & \qw & \qw & \qw & \ctrl{-1} & \qw & \ctrl{-2} & \qw & \qw & \qw & \qw & \qw & \targ & \qw & \qw & \qw & \qw & \qw & \targ & \qw & \qw & \qw & \qw & \qw & \qw & \qw}
}
\caption{Quantum circuit for encoding Steane [[7,1,3]] quantum error correction codeword.\label{fig:encoding}}
\end{figure*}

\begin{figure*}
\mbox{
\centering
\Qcircuit @C=1em @R=.15em @!R {
  & \qw\ar@{--}[]+<0em,.3em>;[dddddd]+<0em,-.3em> & \gate{I} & \gate{E}& \qw \\
  & \qw & \gate{I} & \gate{E}& \qw\\
  & \qw & \gate{I} & \gate{E}& \qw\\
  & \qw & \gate{I} & \gate{E}& \qw\\
  & \qw & \gate{I} & \gate{E}& \qw\\
  & \qw & \gate{I} & \gate{E}& \qw\\
  & \qw & \gate{I} & \gate{E}& \qw}
}
\caption{Quantum circuit for logical identity operation, referred to as a no-op.\label{fig:nop}}
\end{figure*}
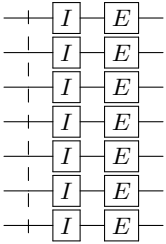

We present the quantum circuits used for the numerical results presented in Sec.~\ref{sec: Steane results}.
This includes the encoding and error correction subcircuits for the Steane [[7,1,3]] code \cite{Steane1996}, which corresponds to the functional blocks shown in Fig.~\ref{fig:circuit overview}.
The first two blocks in that figure correspond to the subcircuits shown in Figs.~\ref{fig:encoding} and \ref{fig:nop}, respectively.
The block ``Error Correction'' in  Fig.~\ref{fig:circuit overview} consists of the six syndrome extraction circuits, which correspond to Figs.~\ref{fig:ext1}--\ref{fig:ext6} below. Syndrome extraction is followed by a recovery operation (not shown) prescribed by the measurement results.
In the case of noisy error correction, the syndrome extraction circuit is repeated three times; a correction is applied only if at least two of syndromes are the same.
In these diagrams, $MR$ denotes a measurement in the computational basis followed by a reset of the qubit to the $\ket{0}$ state.
Dashed vertical lines denote the boundaries at which simulated errors occur.
At such times, each qubit is independently subjected to either amplitude damping or depolarizing error, depending on the noise model chosen for the simulation.
\newcommand{\qwu}{\ar@{--}[]+<0em,.45em>;[d]+<0em,.75em>}

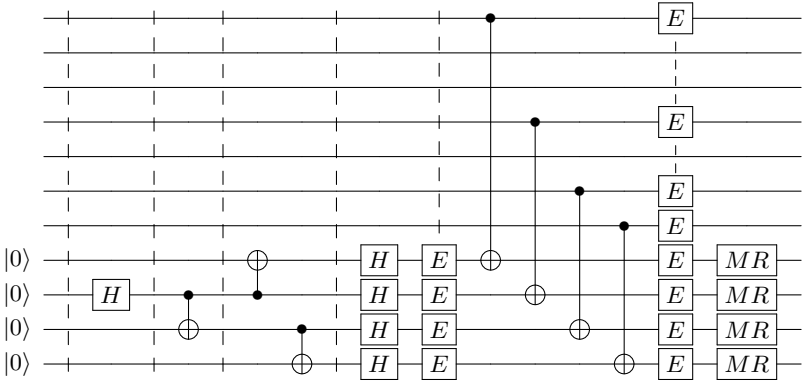
\begin{figure*}
\mbox{
\centering
\Qcircuit @C=1em @R=.15em @!R {
  & \qw\ar@{--}[]+<0em,.3em>;[dddddddddd]+<0em,-.3em> & \qw & \qw\ar@{--}[]+<0em,.3em>;[dddddddddd]+<0em,-.3em> & \qw & \qw\ar@{--}[]+<0em,.3em>;[dddddddddd]+<0em,-.3em> & \qw & \qw & \qw\ar@{--}[]+<0em,.3em>;[dddddddddd]+<0em,-.3em> & \qw &  \qw\ar@{--}[]+<0em,.3em>;[dddddd]+<0em,-.3em> & \ctrl{7} & \qw & \qw & \qw & \gate{E} & \qw & \qw\\
  & \qw & \qw & \qw & \qw & \qw & \qw & \qw & \qw & \qw 
  & \qw & \qw & \qw & \qw & \qw & \qw \qwu & \qw & \qw\\
  & \qw & \qw & \qw & \qw & \qw & \qw & \qw & \qw & \qw 
  & \qw & \qw & \qw & \qw & \qw & \qw \qwu & \qw & \qw\\
  & \qw & \qw & \qw & \qw & \qw & \qw & \qw & \qw & \qw
  & \qw & \qw & \ctrl{5} & \qw & \qw & \gate{E} & \qw & \qw\\
  & \qw & \qw & \qw & \qw & \qw & \qw & \qw & \qw & \qw
  & \qw & \qw & \qw & \qw & \qw & \qw \qwu & \qw & \qw\\
  & \qw & \qw & \qw & \qw & \qw &  \qw & \qw & \qw & \qw
  & \qw & \qw & \qw & \ctrl{4} & \qw & \gate{E} & \qw & \qw\\
  & \qw & \qw & \qw & \qw & \qw &  \qw & \qw & \qw & \qw
  & \qw & \qw & \qw & \qw & \ctrl{4} & \gate{E}& \qw & \qw\\
 \lstick{\ket{0}} & \qw & \qw & \qw & \qw & \qw & \targ & \qw & \qw & \gate{H} & \gate{E} & \targ & \qw & \qw & \qw & \gate{E} & \gate{MR} & \qw\\
 \lstick{\ket{0}} & \qw & \gate{H} & \qw & \ctrl{1} & \qw & \ctrl{-1} & \qw & \qw & \gate{H} & \gate{E} & \qw & \targ & \qw & \qw & \gate{E} & \gate{MR} & \qw\\
 \lstick{\ket{0}} & \qw & \qw & \qw & \targ & \qw & \qw & \ctrl{1} & \qw & \gate{H}  & \gate{E}& \qw & \qw & \targ & \qw & \gate{E} & \gate{MR} & \qw\\
 \lstick{\ket{0}}  & \qw & \qw & \qw & \qw & \qw & \qw & \targ & \qw & \gate{H}  & \gate{E} & \qw & \qw & \qw & \targ & \gate{E} & \gate{MR} & \qw}
}
\caption{Quantum circuit for extracting syndrome 1.\label{fig:ext1}}
\end{figure*}

\begin{figure*}
\mbox{
\centering
\Qcircuit @C=1em @R=.15em @!R {
  & \qw\ar@{--}[]+<0em,.3em>;[dddddddddd]+<0em,-.3em> & \qw & \qw\ar@{--}[]+<0em,.3em>;[dddddddddd]+<0em,-.3em> & \qw & \qw\ar@{--}[]+<0em,.3em>;[dddddddddd]+<0em,-.3em> & \qw & \qw & \qw\ar@{--}[]+<0em,.3em>;[dddddddddd]+<0em,-.3em> & \qw & \qw\ar@{--}[]+<0em,.3em>;[dddddd]+<0em,-.3em> & \qw & \qw & \qw & \qw & \qw \qwu & \qw & \qw \\
  & \qw & \qw & \qw & \qw & \qw & \qw & \qw & \qw & \qw & \qw & \ctrl{6} & \qw & \qw & \qw& \gate{E} & \qw & \qw \\
  & \qw & \qw & \qw & \qw & \qw & \qw & \qw & \qw & \qw & \qw & \qw & \qw & \qw & \qw & \qw \qwu & \qw & \qw \\
  & \qw & \qw & \qw & \qw & \qw & \qw & \qw & \qw & \qw & \qw & \qw & \ctrl{5} & \qw & \qw & \gate{E} & \qw & \qw \\
  & \qw & \qw & \qw & \qw & \qw & \qw & \qw & \qw & \qw & \qw & \qw & \qw & \ctrl{5} & \qw & \gate{E} & \qw & \qw \\
  & \qw & \qw & \qw & \qw & \qw & \qw & \qw & \qw & \qw & \qw & \qw & \qw & \qw & \qw & \qw \qwu & \qw & \qw \\
  & \qw & \qw & \qw & \qw & \qw & \qw & \qw & \qw & \qw & \qw & \qw & \qw & \qw & \ctrl{4} & \gate{E} & \qw & \qw \\
 \lstick{\ket{0}} & \qw & \qw & \qw & \qw & \qw & \targ & \qw & \qw & \gate{H} & \gate{E} & \targ & \qw & \qw & \qw & \gate{E} & \gate{MR} & \qw \\
 \lstick{\ket{0}} & \qw & \gate{H} & \qw & \ctrl{1} & \qw & \ctrl{-1} & \qw & \qw & \gate{H} & \gate{E} & \qw & \targ & \qw & \qw & \gate{E} & \gate{MR} & \qw \\
 \lstick{\ket{0}} & \qw & \qw & \qw & \targ & \qw & \qw & \ctrl{1} & \qw & \gate{H} & \gate{E}& \qw & \qw & \targ & \qw &\gate{E} & \gate{MR} & \qw\\
 \lstick{\ket{0}} & \qw & \qw & \qw & \qw & \qw & \qw & \targ & \qw & \gate{H} & \gate{E} & \qw & \qw & \qw & \targ & \gate{E} & \gate{MR} & \qw}
}
\caption{Quantum circuit for extracting syndrome 2.\label{fig:ext2}}
\end{figure*}
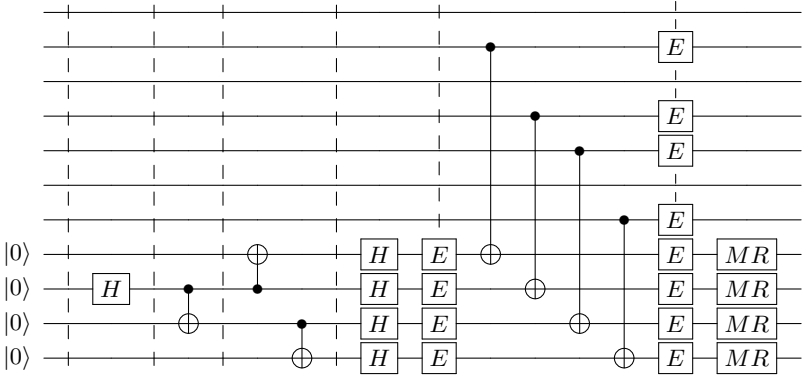

\begin{figure*}
\mbox{
\centering
\Qcircuit @C=1em @R=.15em @!R {
  & \qw\ar@{--}[]+<0em,.3em>;[dddddddddd]+<0em,-.3em> & \qw & \qw\ar@{--}[]+<0em,.3em>;[dddddddddd]+<0em,-.3em> & \qw & \qw\ar@{--}[]+<0em,.3em>;[dddddddddd]+<0em,-.3em> & \qw & \qw & \qw\ar@{--}[]+<0em,.3em>;[dddddddddd]+<0em,-.3em> & \qw & \qw\ar@{--}[]+<0em,.3em>;[dddddd]+<0em,-.3em> & \qw & \qw & \qw & \qw & \qw \qwu & \qw & \qw \\
  & \qw & \qw & \qw & \qw & \qw & \qw & \qw & \qw & \qw & \qw & \qw & \qw & \qw & \qw & \qw  \qwu  & \qw & \qw \\
  & \qw & \qw & \qw & \qw & \qw & \qw & \qw & \qw & \qw & \qw & \ctrl{5} & \qw & \qw & \qw & \gate{E} & \qw & \qw \\
  & \qw & \qw & \qw & \qw & \qw & \qw & \qw & \qw & \qw & \qw & \qw & \ctrl{5} & \qw & \qw & \gate{E} & \qw & \qw \\
  & \qw & \qw & \qw & \qw & \qw & \qw & \qw & \qw & \qw & \qw & \qw & \qw & \ctrl{5} & \qw & \gate{E} & \qw & \qw \\
  & \qw & \qw & \qw & \qw & \qw & \qw & \qw & \qw & \qw & \qw & \qw & \qw & \qw & \ctrl{5} & \gate{E} & \qw & \qw \\
  & \qw & \qw & \qw & \qw & \qw & \qw & \qw & \qw & \qw & \qw & \qw & \qw & \qw & \qw & \qw \qwu & \qw & \qw \\
\lstick{\ket{0}}  & \qw & \qw & \qw & \qw & \qw & \targ & \qw & \qw & \gate{H} & \gate{E} & \targ & \qw & \qw & \qw & \gate{E} & \gate{MR} & \qw \\
\lstick{\ket{0}}  & \qw & \gate{H} & \qw & \ctrl{1} & \qw & \ctrl{-1} & \qw & \qw & \gate{H} & \gate{E} & \qw & \targ & \qw & \qw & \gate{E} & \gate{MR} & \qw\\
\lstick{\ket{0}}  & \qw & \qw & \qw & \targ & \qw & \qw & \ctrl{1} & \qw & \gate{H} & \gate{E} & \qw & \qw & \targ & \qw & \gate{E} & \gate{MR} & \qw \\
\lstick{\ket{0}}  & \qw & \qw & \qw & \qw & \qw & \qw & \targ & \qw & \gate{H} & \gate{E} & \qw & \qw & \qw & \targ & \gate{E} & \gate{MR} & \qw}
}
\caption{Quantum circuit for extracting  syndrome 3.\label{fig:ext3}}
\end{figure*}
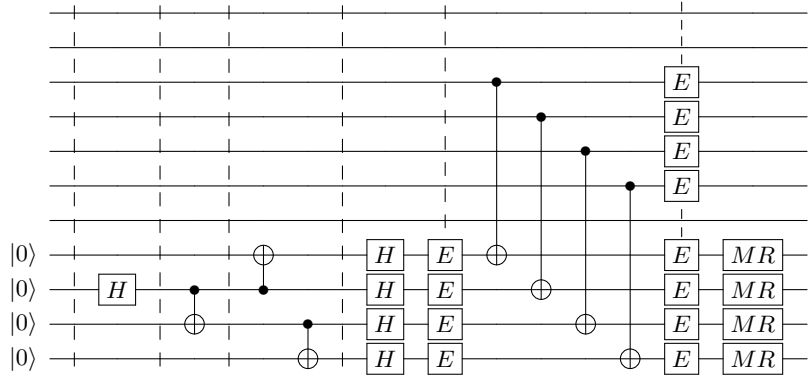

\begin{figure*}
\mbox{\centering
\Qcircuit @C=1em @R=.15em @!R {
  & \qw\ar@{--}[]+<0em,.3em>;[dddddddddd]+<0em,-.3em> & \qw & \qw\ar@{--}[]+<0em,.3em>;[dddddddddd]+<0em,-.3em> & \qw & \qw\ar@{--}[]+<0em,.3em>;[dddddddddd]+<0em,-.3em> & \qw & \qw & \qw\ar@{--}[]+<0em,.3em>;[dddddddddd]+<0em,-.3em> & \targ & \qw & \qw & \qw &  \gate{E} & \qw & \qw\ar@{--}[]+<0em,.3em>;[dddddd]+<0em,-.3em> & \qw & \qw \\
  & \qw & \qw & \qw & \qw & \qw & \qw & \qw & \qw & \qw & \qw & \qw & \qw & \qw \qwu & \qw & \qw & \qw & \qw \\
  & \qw & \qw & \qw & \qw & \qw & \qw & \qw & \qw & \qw & \qw & \qw & \qw & \qw \qwu & \qw & \qw & \qw & \qw \\
  & \qw & \qw & \qw & \qw & \qw & \qw & \qw & \qw & \qw & \targ & \qw & \qw & \gate{E} & \qw & \qw & \qw & \qw \\
  & \qw & \qw & \qw & \qw & \qw & \qw & \qw & \qw & \qw & \qw & \qw & \qw & \qw \qwu & \qw & \qw & \qw & \qw \\
  & \qw & \qw & \qw & \qw & \qw & \qw & \qw & \qw & \qw & \qw & \targ & \qw & \gate{E} & \qw & \qw & \qw & \qw \\ 
  & \qw & \qw & \qw & \qw & \qw & \qw & \qw & \qw & \qw & \qw & \qw & \targ & \gate{E}& \qw & \qw & \qw & \qw \\ 
\lstick{\ket{0}}  & \qw & \qw & \qw & \qw & \qw & \targ & \qw & \qw & \ctrl{-7} & \qw & \qw & \qw & \gate{E} & \gate{H} & \gate{E}& \gate{MR} & \qw\\
\lstick{\ket{0}}  & \qw & \gate{H} & \qw & \ctrl{1} & \qw & \ctrl{-1} & \qw & \qw & \qw & \ctrl{-5} & \qw & \qw & \gate{E}& \gate{H} & \gate{E} & \gate{MR} & \qw\\
\lstick{\ket{0}}  & \qw & \qw & \qw & \targ & \qw & \qw & \ctrl{1} & \qw & \qw & \qw & \ctrl{-4} & \qw & \gate{E}& \gate{H} & \gate{E} & \gate{MR} & \qw\\
\lstick{\ket{0}}  & \qw & \qw & \qw & \qw & \qw & \qw & \targ & \qw & \qw & \qw & \qw & \ctrl{-4} & \gate{E} & \gate{H} & \gate{E} & \gate{MR} & \qw}
}
\caption{Quantum circuit for extracting  syndrome 4.\label{fig:ext4}}
\end{figure*}
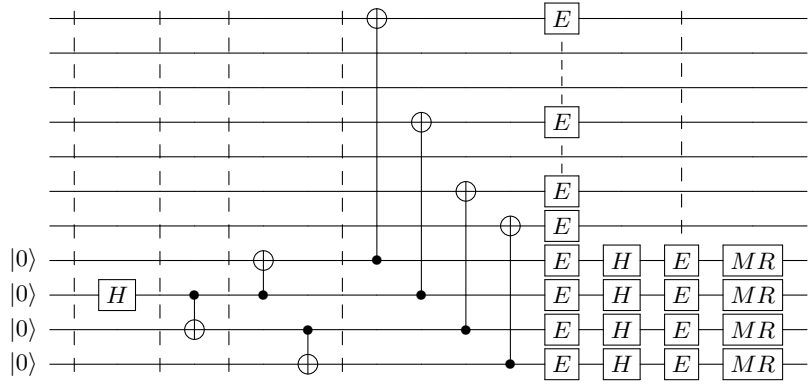

\begin{figure*}
\mbox{\centering
\Qcircuit @C=1em @R=.15em @!R {
  & \qw\ar@{--}[]+<0em,.3em>;[dddddddddd]+<0em,-.3em> & \qw & \qw\ar@{--}[]+<0em,.3em>;[dddddddddd]+<0em,-.3em> & \qw & \qw\ar@{--}[]+<0em,.3em>;[dddddddddd]+<0em,-.3em> & \qw & \qw & \qw\ar@{--}[]+<0em,.3em>;[dddddddddd]+<0em,-.3em> & \qw & \qw & \qw & \qw & \qw \qwu  & \qw & \qw\ar@{--}[]+<0em,.3em>;[dddddd]+<0em,-.3em> & \qw & \qw\\
  & \qw & \qw & \qw & \qw & \qw & \qw & \qw & \qw & \targ & \qw & \qw & \qw & \gate{E} & \qw & \qw & \qw & \qw \\
  & \qw & \qw & \qw & \qw & \qw & \qw & \qw & \qw & \qw & \qw & \qw & \qw & \qw \qwu & \qw & \qw & \qw & \qw \\
  & \qw & \qw & \qw & \qw & \qw & \qw & \qw & \qw & \qw & \targ & \qw & \qw & \gate{E} & \qw & \qw & \qw & \qw \\
  & \qw & \qw & \qw & \qw & \qw & \qw & \qw & \qw & \qw & \qw & \targ & \qw & \gate{E} & \qw & \qw & \qw & \qw \\
  & \qw & \qw & \qw & \qw & \qw & \qw & \qw & \qw & \qw & \qw & \qw & \qw & \qw \qwu & \qw & \qw & \qw & \qw \\
  & \qw & \qw & \qw & \qw & \qw & \qw & \qw & \qw & \qw & \qw & \qw & \targ & \gate{E} & \qw & \qw & \qw & \qw \\
 \lstick{\ket{0}} & \qw & \qw & \qw & \qw & \qw & \targ & \qw & \qw & \ctrl{-6} & \qw & \qw & \qw & \gate{E} & \gate{H} & \gate{E} & \gate{MR} & \qw\\
\lstick{\ket{0}}  & \qw & \gate{H} & \qw & \ctrl{1} & \qw & \ctrl{-1} & \qw & \qw & \qw & \ctrl{-5} & \qw & \qw & \gate{E} & \gate{H} & \gate{E} & \gate{MR} & \qw\\
\lstick{\ket{0}} & \qw & \qw & \qw & \targ & \qw & \qw & \ctrl{1} & \qw & \qw & \qw & \ctrl{-5} & \qw & \gate{E} & \gate{H} & \gate{E} & \gate{MR} & \qw\\
\lstick{\ket{0}}  & \qw & \qw & \qw & \qw & \qw & \qw & \targ & \qw & \qw & \qw & \qw & \ctrl{-4} & \gate{E} & \gate{H} & \gate{E} & \gate{MR} & \qw}
}
\caption{Quantum circuit for extracting syndrome 5.\label{fig:ext5}}
\end{figure*}
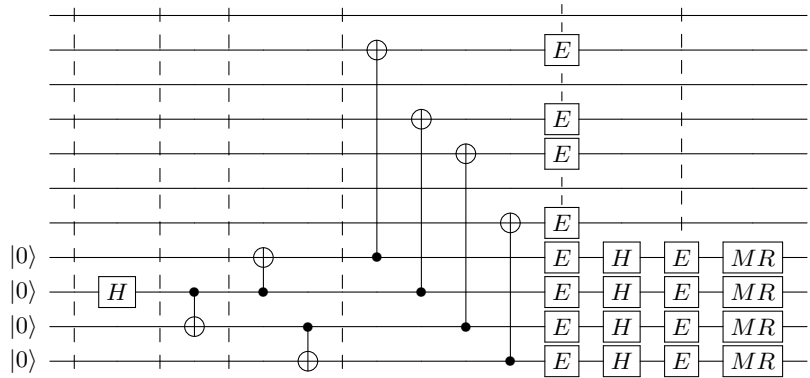

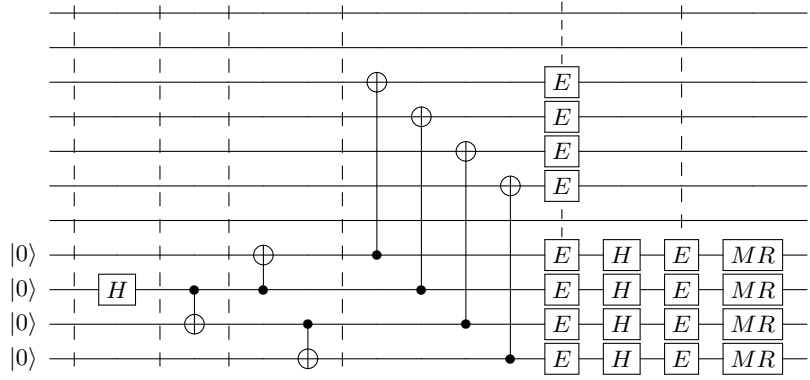
\begin{figure*}
\mbox{\centering
\Qcircuit @C=1em @R=.15em @!R {
  & \qw\ar@{--}[]+<0em,.3em>;[dddddddddd]+<0em,-.3em> & \qw & \qw\ar@{--}[]+<0em,.3em>;[dddddddddd]+<0em,-.3em> & \qw & \qw\ar@{--}[]+<0em,.3em>;[dddddddddd]+<0em,-.3em> & \qw & \qw & \qw\ar@{--}[]+<0em,.3em>;[dddddddddd]+<0em,-.3em> & \qw & \qw & \qw & \qw & \qw \qwu  & \qw & \qw\ar@{--}[]+<0em,.3em>;[dddddd]+<0em,-.3em> & \qw & \qw\\
  & \qw & \qw & \qw & \qw & \qw & \qw & \qw & \qw & \qw & \qw & \qw & \qw & \qw \qwu & \qw & \qw & \qw & \qw \\
  & \qw & \qw & \qw & \qw & \qw & \qw & \qw & \qw & \targ & \qw & \qw & \qw & \gate{E} & \qw & \qw & \qw & \qw \\
  & \qw & \qw & \qw & \qw & \qw & \qw & \qw & \qw & \qw & \targ & \qw & \qw & \gate{E} & \qw & \qw & \qw & \qw \\
  & \qw & \qw & \qw & \qw & \qw & \qw & \qw & \qw & \qw & \qw & \targ & \qw & \gate{E} & \qw & \qw & \qw & \qw \\
  & \qw & \qw & \qw & \qw & \qw & \qw & \qw & \qw & \qw & \qw & \qw & \targ & \gate{E} & \qw & \qw & \qw & \qw \\
  & \qw & \qw & \qw & \qw & \qw & \qw & \qw & \qw & \qw & \qw & \qw & \qw & \qw \qwu  & \qw & \qw & \qw & \qw \\
\lstick{\ket{0}}  & \qw & \qw & \qw & \qw & \qw & \targ & \qw & \qw & \ctrl{-5} & \qw & \qw & \qw & \gate{E} & \gate{H} & \gate{E} & \gate{MR} & \qw\\
\lstick{\ket{0}}  & \qw & \gate{H} & \qw & \ctrl{1} & \qw & \ctrl{-1} & \qw & \qw & \qw & \ctrl{-5} & \qw & \qw & \gate{E} & \gate{H} & \gate{E} & \gate{MR} & \qw\\
\lstick{\ket{0}}  & \qw & \qw & \qw & \targ & \qw & \qw & \ctrl{1} & \qw & \qw & \qw & \ctrl{-5} & \qw & \gate{E}& \gate{H} & \gate{E} & \gate{MR} & \qw\\
\lstick{\ket{0}}  & \qw & \qw & \qw & \qw & \qw & \qw & \targ & \qw & \qw & \qw & \qw & \ctrl{-5} & \gate{E} & \gate{H} & \gate{E}& \gate{MR} & \qw}
}
\caption{Quantum circuit for extracting syndrome 6.\label{fig:ext6}}
\end{figure*}

\end{document}